\documentstyle[amssymb,prl,aps,psfig]{revtex}

\begin{document}
\title{Suppression and enhancement of the critical current in multiterminal S/N/S
mesoscopic structures.}
\author{R. Seviour$^{\ast }$ and A.F. Volkov$^{\ast \dagger }$ .}
\address{$^*$ School of Physics and Chemistry,\\
Lancaster University, Lancaster LA1 4YB, U.K.\\
$^{\dagger}$Institute of Radioengineering and Electronics of the Russian \\
Academy of Sciencies, Mokhovaya str.11, Moscow 103907, Russia.}
\date{\today}
\maketitle

\begin{abstract}
We analyse the measured critical current $I_{m\text{ }}$ in a mesoscopic
4-terminal S/N/S structure. The current through the S/N interface is shown
to consist not only of the Josephson component $I_{c}\sin \varphi ,$ but
also a phase-coherent part $I_{sg}\cos \varphi $ of the subgap current. The
current $I_{m}$ is determined by the both components $I_{c}$ and $I_{sg},$
and depends in a nonmonotonic way on the voltage $V$ between superconductors
and normal reservoirs reaching a maximum at $V\cong \Delta /e$. The obtained
theoretical resultas are in qualitative agreement with recent experimental
data.
\end{abstract}

Recent achievements in nanotechnology have revived\smallskip\ interest in
the study of nonequilibrium and phase-coherent phenomena in
superconductor-normal metal (S/N) structures. One of the most remarkable,
discovered recently \cite{r1}, was the observation of the sign reversal of
the Josephson critical current $I_{c}$ (the so-called $\pi $-junction) in a
multi-terminal mesoscopic Nb/Au/Nb structure under nonequilibrium conditions.
By passing an additional
current through the N layer or, in another words, by applying a voltage $V$
to the normal reservoirs (see Fig.1) with respect to the superconductors,
one can create a nonequlibrium electron-hole distribution, or at least one can
shift this distribution with respect to the electron-hole distribution 
in the superconductors. Under this
condition, the critical current $I_{c}$ decreases with $V$ and changes sign
at a certain value of the applied voltage $V$. This effect was predicted
first in Ref. \cite{r6} where a ballistic 3-terminal structure was
considered (for more details, see also Refs.\cite{r7,r8}). In diffusive
4-terminal S/N/S structures, the sign-reversal effect has been considered in
Refs.\cite{r9,r10,r11} (see also \cite{r12,r13}). The
sign-reversal effect and switching of the $\pi $-junction into a state where 
$\varphi =\pi $ has much in common with an instability of an uniform
superconductor with a nonequilibrium distribution function \cite{r14,r15}.

In multi-terminal S/N/S structures one can observe not only the sign
reversal effect, but also a number of other interesting phenomena. For
example, the conductance of a normal wire between N reservoirs oscillates
with varying phase difference $\varphi $ (see review articles \cite{r16,r17}%
). In addition, as shown in Refs. \cite{r9,r18}, the measured critical
current $I_{m}$ depends on the geometry of a particular structure and
instead of decreasing may also increase with increasing voltage $V$. In
particular one can observe Josephson-like effects (plateau on the $%
I_{3}(V_{S})$ curve, oscillations of the measured critical current $I_{m}$
in a magnetic field etc) even if the Josephson coupling between
superconductors under equilibrium conditions is negligable.  The reason for 
these effects is that the current $I_{m}$ in a multi-terminal S/N/S structure 
is determined not only by the Josephson component $I_{c}\sin \varphi$, but also 
by the phase-dependent subgap current $I_{sg}\cos \varphi$ through the S/N interface. 
Therefore even in the case of a small $I_{c}$, the current $I_{m}$ can be altered 
by varying the phase $\phi$. An increase of
the critical current was observed in the recent paper \cite{r19} where a
mesoscopic three-terminal S/N/S structure was studied. The authors used a
third superconductor as a reservoir the electric potential of which was
shifted with respect to the other two superconductors by the voltage $V$.
The measured critical current reached its maximal value when the magnitude
of $V$ was comparable with $\Delta .$ At some, not too low temperatures $T$
the measured critical current $I_{m}$ exceeds its magnitude in the
equilibrium state: $I_{m}(V)>I_{m}(0)$.
In the present paper we show that the enhancement of the supercurrent
observed in Ref.\cite{r19} is most likely caused by the mechanism mentioned above. 
In Refs.\cite{r9,r18} the model case of gapless
superconductors was considered where there is no singularity in the
density-of states in superconductors at $\epsilon =\Delta .$ Here we will
consider the case of ordinary superconductors with an energy gap $\Delta $
and show that the enhancement of the critical current reaches a maximum for $%
V$ of order $\Delta .$ The voltage dependence $I_{m}$ $(V)$ calculated for
different temperatures is in qualitative agreement with the experimental
data.

We consider the structure shown in\ Fig.1 which differs from the structure
studied experimentally. However in our opinion, this difference is not
essential and allows us to give at least qualitative explanation for the
phenomena observed in Ref.\cite{r19}. First, we assume for simplicity that
the structure under consideration is symmetrical, i.e. it has four terminals
and not three as in the experiment. Secondly, we consider normal reservoirs
in order to avoid complications which would arise in case of superconducting
reservoirs (ac Josephson effects when the finite voltage is applied to the S
reservoir). We also assume for simplicity that the contacts between the N
wire and N reservoirs are good (the resistance of the N wire/N reservoir interface is
much smaller than the resistance of the N wire), whereas the S/N interface
resistance is finite (larger or less than the resistance of the N wire). We
will study the diffusive case which corresponds to the experiment \cite{r19}.

In order to find the dependence of the effective critical current $I_{m}(V)$
( the definition of $I_{m}(V)$ will be given later), we need to determine
two distribution function $f_{+}$ and $f_{-}$. Both these functions are
isotropic in space. The function $f_{+}$ is related to a symmetrical part of
the distribution function in the electron-hole space:$f_{+}(\epsilon
)=1-(n_{\uparrow }(\epsilon )+p_{\downarrow }(\epsilon ))=1-(n_{\downarrow
}(\epsilon )+p_{\uparrow }(\epsilon ))$, here $p_{\downarrow }(\epsilon
)=1-n_{\downarrow }(-\epsilon )$ is the hole distribution function. It
determines the critical current $I_{c}.$ The function $f$ $_{-}$\ describes
the electron-hole imbalance and determines the electric potential and
current: $f(\epsilon )=-(n_{\uparrow }(\epsilon )-p_{\downarrow }(\epsilon
))=-(n_{\downarrow }(\epsilon )-p_{\uparrow }(\epsilon ))$. Equations for $f_{+}$
and $f_{-}$ are obtained from an equation for the matrix Keldysh function $%
\stackrel{\wedge }{G}$ (see, for example \cite{r9,r17}  For the
structure shown in\ Fig.1 they can be written in the form

\begin{equation}
L{\Large \partial }_{x}{\Large (M}_{-}{\Large \partial }_{x}f_{-}(x){\Large %
+J}_{s}{\Large f}_{+}{\Large -J}_{an}{\Large \partial }_{x}f_{+}(x){\Large %
)=r[}A_{-}{\Large \delta (x-L_{1})+}\stackrel{\_}{A_{-}}{\Large \delta (x+L_{1})].}
\end{equation}

\begin{equation}
L{\Large \partial }_{x}{\Large (M}_{+}{\Large \partial }_{x}f_{+}(x){\Large %
+J}_{s}{\Large f}_{-}{\Large +J}_{an}{\Large \partial }_{x}f_{-}(x){\Large %
)=r[}A_{+}{\Large \delta (x-L_{1})+}\stackrel{\_}{A_{+}}{\Large \delta (x+L_{1})]}
\end{equation}

Here all the coefficients are expressed through the retarded (advanced)
Green's functions $\stackrel{\wedge }{G^{R}}=G^{R}\hat{\sigma}_{z}+\stackrel{%
\wedge }{F^{R}}$ and are equal to $M_{\pm}=(1-G^{R}G^{A}\mp (\stackrel{%
\wedge }{F^{R}}\stackrel{\wedge }{F^{A}})_{1})/2;$

$J_{an}=\stackrel{\wedge }{(F^{R}}\stackrel{\wedge }{F^{A}})_{z}/2,\
J_{s}=(1/2)(\stackrel{\wedge }{F^{R}}\stackrel{\wedge }{\partial _{x}F^{R}}-%
\stackrel{\wedge }{F^{A}}\stackrel{\wedge }{\partial _{x}F^{A}})_{z},$\ $%
A_{-}=(\nu \nu _{S}+g_{1+})f_{-}-(g_{z-}f_{eq}+g_{z+}f_{+});$ $A_{+}=(\nu
\nu +g_{1-})(f_{+}-f_{eq})-g_{z-}f_{-};g_{1\pm }=(1/4)[(\stackrel{\wedge }{%
F^{R}}\stackrel{\wedge }{\pm F^{A}})(\stackrel{\wedge }{F_{S}^{R}}\pm 
\stackrel{\wedge }{F_{S}^{A}})]_{1};$ $g_{z\pm }=(1/4)[(\stackrel{\wedge }{%
F^{R}}\stackrel{\wedge }{\mp F^{A}})(\stackrel{\wedge }{F_{S}^{R}}\pm 
\stackrel{\wedge }{F_{S}^{A}})]_{z};$

The coefficient $r=R/R_{b}$, $R=\rho L/d$ is the resistance of the N film per unit
length in the z-direction,$\rho $ is the specific resistivity of the N film, 
$d$ is the thickness of the N film, $R_{b}$ is the S/N interface resistance;
the functions $\stackrel{\_}{A_{-}}$ and $\stackrel{\_}{A}_{+}$ coincide with $A_{-}$
and $A_{+}$ if we make a substitution $\varphi \rightarrow -\varphi $. We
introduced above the following notations $(\stackrel{\wedge }{F^{R}}%
\stackrel{\wedge }{F^{A}})_{1}=Tr(\stackrel{\wedge }{F^{R}}\stackrel{\wedge 
}{F^{A}})/2,$ $(\stackrel{\wedge }{F^{R}}\stackrel{\wedge }{F^{A}})_{z}=Tr(%
\hat{\sigma}_{z}\stackrel{\wedge }{F^{R}}\stackrel{\wedge }{F^{A}})/2$ etc.; 
$\nu $ and $\nu _{S}$ are the density-of states in the N film at $x=L_{1}$
and in the superconductors. The boundary conditions for $f_{+\text{ }}$and $%
f_{-}$ are: $f_{+\text{ }}(L)=F_{V+}$ and $f_{-}(L)=F_{V-}$; the functions $%
F_{V\pm }$ are the corresponding distribution functions in the normal
reservoirs: $F_{V\pm }=[\tanh ((\epsilon +eV)\beta )\pm \tanh ((\epsilon
-eV)\beta )]/2$. We set the electrical potential at the superconductors
equal to zero and assumed that the width of the S/N interfaces $w$ is small
compared to $L_{1,2}$.

 Eq.(1) describes the conservation of the electric current (at a given energy). 
The term in the brackets on the left is the total partial current in the N wire,
consisting of the quasiparticle current (the first term), the supercurrent in 
the interval $(-L_{1},L_{1})$ (the second term) and a "nonequilibrium supercurrent"
(the third term). The coefficient $M$ is a quantity which is proportional to the diffusion 
coefficient renormalised due to proximity effect. The right hand side is 
the partial current through the S/N interface; the term $(\nu \nu _{S}+g_{1+})f_{-}$
is the quasiparticle current above ($\nu \nu _{S}f_{-}$) and below ($g_{1+}f_{-}$) the gap.
The term ($g_{z-}f_{eq}+g_{z+}f_{+}$) is the Josephson current in nonequilibrium conditions.
Eq.(2) describes the conservation of the energy flux (at a given energy). The 
coefficient $A_{+}$ is zero below the gap ( complete Andreev reflection) as the difference
($F_{S}^{R} - F_{S}^{A}$) equals zero at $\epsilon < \Delta$.

The solutions of Eqs.(1)-(2) can be found exactly and expressed in terms of
the retarded (advanced) Green's functions which obey the Usadel equation.
First we note that the expressions in brackets in the left hand side of
Eqs.(1)-(2) in the regions ($0,L_{1}$) and ($L_{1},L$) are equal to the
constants of integration $C_{1,2 \pm}$. The constants $C_{1,2-}$
relate to partial currents $J_{1,2}$ ($C_{1,2-}=eJ_{1,2}\rho /d)$. The
partial currents $J_{1,2}$ are the currents per unit energy and connected
with the electrical currents $I_{1,2}$ via the relation

\begin{equation}
I_{1,2}=\int_{0}^{\infty }d\epsilon J_{1,2}(\epsilon )
\end{equation}

\bigskip

Our aim is to find the current $I_{3}$ and express it in terms of the
control current $I_{2}$ (or voltage $V$) and the phase difference $\varphi $%
. We note that the distribution functions $f_{\pm}(x)$ are
constants in the region $x\in (0,L_{1})$ and vary in the region $x\in
(L_{1},L)$ reaching $F_{V\pm}$ at $x=L$. Dropping
details of calculations, we present final results for limiting cases.

a) Large interface resistance: $r<<1$.

One can show that in this case $f_{+}(0)\cong (F_{V+}+f_{eq}(r_{2}\nu \nu
_{s}))/(1+r_{2}\nu \nu _{s})$ and $f_{-}(0)\cong F_{V-}/(1+r_{2}\nu \nu _{s})
$, where $r_{2}=r(L_{2}/L).$\ The current $I_{3}$ through the S/N interface
consists of three terms

\begin{equation}
I_{3}(V)=I_{o}(V)-I_{c}(V)\sin \varphi +I_{sg}(V)\cos \varphi 
\end{equation}

\bigskip

Two of them ($I_{o},I_{sg}\cos \varphi $) are the quasiparticle currents and
one ($I_{c}\sin \varphi $) is the Josephson current. This expression shows
that at a given control voltage $V$ and zero voltage difference between the
superconductors ($\varphi $ is constant in time) the current $I_{3}$ may
vary with changing $\varphi $ in the limits:$\mid I_{3}(V)-I_{o}(V)\mid \leq
I_{m}(V)$. This means a plateau on the $V_{S}(I_{3})$ characteristics (see 
\cite{r9,r18}); here $V_{S}=(\hbar /2e)\partial _{t}\varphi $ is the voltage
difference between superconductors. We can write the phase-dependent part of 
$I_{3}$ in the form $I_{3\varphi }=I_{m}\sin (\varphi +\alpha )$, where $%
I_{m}=\sqrt{I_{c}^{2}+I_{sg}^{2}}$ is the measured critical current, $\cos
\alpha =-I_{c}/I_{m}$. In the considered limit of high interface resistance,
we have for $I_{c}$ and $I_{sg}$

$I_{c}(eR_{b})=-\int_{0}^{\infty }d\epsilon \{%
\mathop{\rm Im}%
(F_{S}(F_{y}-F_{x}))f_{o}(0)+%
\mathop{\rm Re}%
F_{S}%
\mathop{\rm Im}%
(F_{y}-F_{x})f(0)\}$

$I_{sg}(eR_{b})=\int_{0}^{\infty }d\epsilon g_{sg}f(0)=\int_{0}^{\infty
}d\epsilon 
\mathop{\rm Im}%
F_{S}%
\mathop{\rm Im}%
(F_{y}-F_{x})f(0);$

Here $\theta =k_{\epsilon }L,\theta _{2}=k_{\epsilon }L_{2},k_{\epsilon }=%
\sqrt{(-2i\epsilon +\gamma )/D},$ $\gamma $ and $D$ is the damping rate and
diffusion coefficient in the N film, $g_{sg}=g_{1+}$ is the normalised
subgap conductance (see the expression for $A$). The functions $F_{y},F_{x}$
are the components of the retarded Green's function in the N film: $%
\stackrel{\wedge }{F^{R}}=F_{x}i\hat{\sigma}_{x}+F_{y}i\hat{\sigma}_{y},$
and $F_{S}=\Delta /\sqrt{(\epsilon +i\Gamma )^{2}-\Delta ^{2}}$ is the
amplitude of the retarded Green's function in the superconductors. If we
linearise the Usadel equation, we obtain $F_{y}-F_{x}=2F_{S}\sinh ^{2}\theta
_{2}/(\theta \sinh 2\theta )$. We note that the numerical solution of the
Usadel equation shows that the linearised solution is a good approximation
even if $r\cong 1$  (at $r=1$ the difference between the exact and the linearized solutions 
at the characteristic energy $\epsilon = \epsilon_{L}=D/L^2$ is less than 5 percent). 
In Fig.2 we plot the $V$ dependence of $I_{c},I_{sg}$
and $I_{m}$ where we see that the real critical current $I_{c}$ decreases
and changes sign with increasing $V$, whereas the measured critical current 
$I_{m}$ first decreases and then increases. Its maximum may exceed $I_{c}(0).$
The reason for such a behaviour of $I_{m}$ is the third term on the right
side in Eq.(5) which describes a contribution of the phase-dependent part of
the subgap quasiparticle current $I_{sg}$ through the S/N interface to the
current $I_{3}$. The current $I_{sg}$ is zero at $V=0$ and increases with $V;
$ this current leads to a low \cite{r21} and high \cite{r22} temperature
peak in the conductance. Its phase dependence was measured in Ref. \cite{r23}
and discussed in many papers (see review articles \cite{r16,r17}). One can
see from Fig.2 that due to the current $I_{sg}$ the measured critical
current $I_{m}$ remains finite when $I_{c}(V)$ turns to zero.

Fig.3 shows the dependence of the measured critical current $I_{m}$ on the
control voltage $V$ for different temperatures. Our results qualitatively
agree with the experimental data of Ref. \cite{r19}; that is, the current $%
I_{m}$ reaches a maximum at $V\cong \Delta /e$ and this maximum exceeds the
equilibrium value of $I_{c}=I_{c}(0)$ when the temperature is not too low. 
 One can see, in agreement with the experimental results of Ref. \cite{r19}, 
the maximal value of $I_{m}$ depends on the temperature much weaker than $I_{c}$.
Although it is difficult to carry out a quantitative comparision between
theory and experiment because in the experiment the width $w$ \ and the
interface resistance $R_{b\text{ }}$were comparable with $L_{1,2}$ and $R$
respectively, and a superconducting reservoir was used instead of a normal
one (therefore, strictly speaking, one must take into account ac Josephson
effects).

An important point to note is that our results do not mean that the sign
reversal of the real critical current $I_{c}$ can not be identified
directly. Consider for example a fork-shape circuit; this means that two
vertical superconducting leads in Fig.1 are attached to a T-shape (inverted)
superconducting lead. Analysing the stability of the state with negative $%
I_{c}$, one can easily show that the state with $\varphi =0$ is unstable
with respect to fluctuations of $\varphi $ and the system switches to a
state with a circulating current. Indeed, taking into account the
fluctuating voltage at the superconductor $V_{S}=\hbar \partial _{t}\varphi
/2e,$ we replace $V$ in Eq.(4) by $V-V_{S}$. We then write down the equation
for the current $\overline{I_{3}}$ in the lead attached to the left
superconductor; this equation coincides with Eq.(4) if $\varphi $ is
replaced by $-\varphi $. Subtracting these equations for $I_{3}$ and $%
\overline{I_{3}}$, we arrive at the equation for a circulating current $%
I_{cir}=-(I_{3}-$ $\overline{I_{3}})/2$:

\begin{equation}
I_{cir}=I_{c}(V)\sin \varphi +V_{S}(R_{0}+R_{sg}\cos \varphi ).
\end{equation}

where $R_{0}=\partial I_{o}/\partial V$ and $R_{sg}=\partial I_{sg}/\partial
V$. Fluctuations of $I_{cir}$ lead to a magnetic flux $\Phi =I_{cir}L/c$ in
the loop which is related to $\varphi $: $\Phi =\Phi _{o}\varphi $, here $%
\Phi _{o}$ is the magnetic flux quantum and we assumed the absence of flux
in the ground state. We find readily from Eq.(5) that the state with $%
\varphi =0$ is unstable if $I_{c}(V)<0$ and $\mid I_{c}(V)\mid >c\Phi _{o}/L,
$ where $L$ is the loop inductance \cite{r24}.

b) Small interface resistance.

One can show that in this case the function $f_{-}(0)$ is zero in the main
approximation with respect to the parameter $(r\theta )^{-1}$ (this means
that the condition $r^{2}>>\Delta /\epsilon _{L}$ should be satisfied; here $%
\epsilon _{L}=D/L^{2}$ is the Thouless energy). The function $f_{+}$ , which
determines the Josephson current, in the main approximation is equal to $%
F_{V+}$ at $\mid \epsilon \mid <\Delta $ and to $f_{eq}$ at $\mid \epsilon
\mid >\Delta $. Therefore the dependence $I_{c}(V)$ is similar to that found
numerically in Ref.\cite{r10} for another geometry (for small interface
resistance); that is, the critical current $I_{c}(V)$ changes sign with
increasing $V$ \ at $V$ of the order of the Thouless energy. As to the
current $I_{2}$, it does not depend on the phase difference in the main
approximation. Indeed, in order to find $I_{2}$ we need to solve the Usadel
equation in the region $x\in (L_{1},L)$ with boundary condition which is
reduced to $\stackrel{\wedge }{G^{R}}=\stackrel{\wedge }{G_{S}^{R}}$. Making the gauge
transformation $\stackrel{\wedge }{G_{S}^{R}}\Longrightarrow $ $\stackrel{%
\wedge }{S}$ $\stackrel{\wedge }{G_{S}^{R}}$ $\stackrel{\wedge }{S^{+}}$, we
can exclude the phase (here $\stackrel{\wedge }{S}=\cos (\varphi /2)+i\hat{%
\sigma}_{z}\sin (\varphi /2)$). Therefore in the main approximation the
third term in Eq.(4) is zero. 

In conclusion, we have studied the dependence of the measured critical
current $I_{m}$ on the voltage $V$ between normal reservoirs and
superconductors in a 4-terminal S/N mesoscopic structure. The current $I_{m}$
is shown to decrease with increasing $V$, then to increase reaching a
maximum at $V\cong \Delta /e.$ Our results qualitatively agree with
experimental data obtained in the recent paper \cite{r19}.

We are grateful to the EPSRC for their financial support.

\begin{figure}
\centerline{\psfig{figure=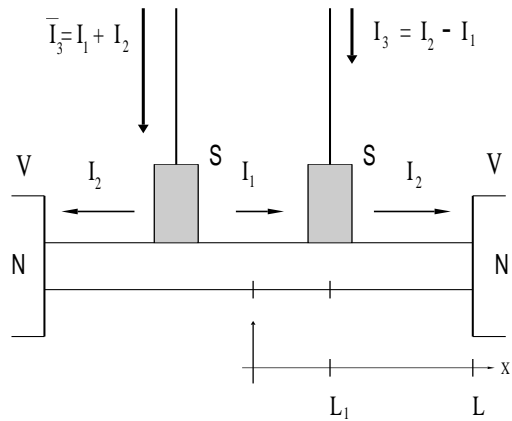,width=8cm,height=6cm}}
\caption{ Schematic view of the 4-terminal S/N/S structure under consideration.
The electric potential of the superconductors is zero.}
\label{fig1}
\end{figure}

\begin{figure}
\centerline{\psfig{figure=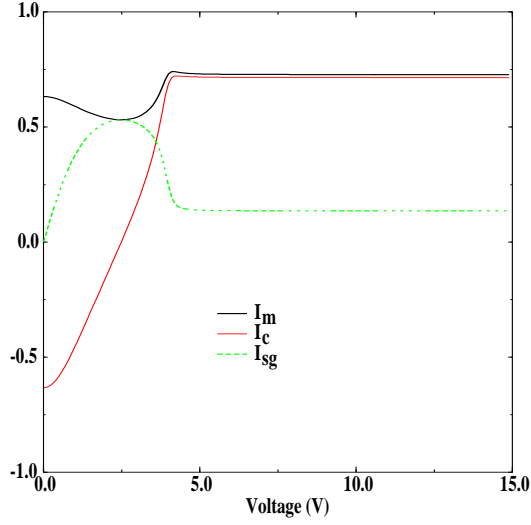,width=8cm,height=8cm}}
\caption{The measured ($I_m$) and real ($I_c$) critical
 currents vs the control voltage $V$. The amplitude of  
 the phase-dependent part ($I_{sg}$) of the subgap current is 
 shown by the dashed line. The currents and voltage are
measured in units $\epsilon _{L}R/eR_{b}^{2}$ and  $\epsilon _{L}/e$
respectively ( $\epsilon _{L}=\hbar D/L^{2}$ is the Thouless energy). The
parameters are: $\Delta =4\epsilon _{L}$,$T=\epsilon _{L}/4$,$L_{1}/L=0.3, r=0.3$.}
\label{fig2}
\end{figure}

\begin{figure}
\centerline{\psfig{figure=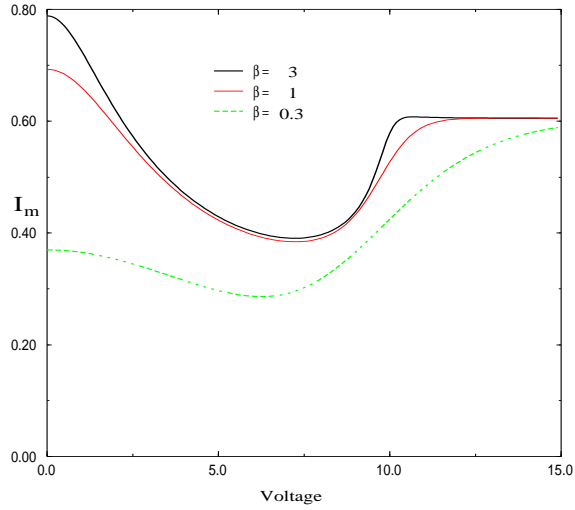,width=8cm,height=8cm}}
\caption{ The measured critical current ($I_{m}$) vs $V$ for different
temperatures: $\beta =\epsilon _{L}/2T$. The parameters are: $\Delta
=10\epsilon _{L},L_{1}/L=0.3, r=0.3$.}
\label{fig3}
\end{figure}
\end{document}